\documentclass[twocolumn,nofootinbib,showpacs,prd]{revtex4-1}%
\usepackage{amsmath}
\usepackage{amsfonts}
\usepackage{amssymb}
\usepackage{graphicx}                

\begin{document}

\title{Second order invariants and holography}

\author{
Luca Bonanno$^\P{}^\flat$, Gerardo Iannone${}^\flat$ and Orlando
Luongo${}^\ddag{}^\dag{}^\S$}

\address{${}^\ddag$Dip. di Fisica, Universit\`a di Roma "La Sapienza", I-00185 Roma, Italy,}

\address{${}^\flat$Institut f$\ddot{u}$r Theoretische Physik, J. W. Goethe - Univ. Max-von-Laue-Stra${\ss}$e, 1 60438
Frankfurt am Main, Germany,}

\address{${}^\P$Dip. di Fisica, Universit\`a di Ferrara and INFN, via Saragat 1, 44100 Ferrara, Italy,}

\address{${}^\dag$Dip. di Scienze Fisiche, Universit\`a di Napoli ''Federico II'', Via Cinthia, Napoli, Italy,}

\address{${}^\S$Instituto de Ciencias Nucleares, Universidad Nacional Autonoma de M\'exico,}

\address{${}^\flat$ Dip. di Fisica, Universit\`a di Salerno and INFN,I-84081, Salerno,
Italy.}

\begin{abstract}
Motivated by recent works on the role of the Holographic principle
in cosmology, we relate a class of
second order Ricci invariants to the IR cutoff characterizing the
holographic Dark Energy density. The choice of second order invariants provides an
invariant way to account the problem of causality for the correct
cosmological cutoff, since the presence of event horizons is not an
\emph{a priori} assumption. We find that these models work fairly
well, by fitting the observational data, through a combined
cosmological test with the use of SNeIa, BAO and CMB. This class of
models is also able to overcome the fine-tuning and coincidence
problems. Finally, to make a comparison with other recent models, we
adopt the statistical tests AIC and BIC.
\end{abstract}


\maketitle

\section{Introduction}

Ever since the discovery of the Dark Energy (DE)
\cite{SNeIa,SN2,sn3}, the fundamental problem of understanding the
nature of the positive acceleration of the universe appears lacking
of some ingredients in the framework of General Relativity (GR). In
addition, observations strongly suggest that matter in the universe
is dominated by a non-baryonic (cold) dark matter (DM) which seems
to evolve separately from the unknown force driving the acceleration
\cite{tr1,tr2}. Despite neither DE nor DM have been directly
detected, they represent the most relevant amount of energy and
matter in the universe, being almost the 70 and 25 percent of the
total content, respectively. A theory propounded to explain their
effects and to describe their natures is strictly necessary in cosmology, 
although, up to now, it has not been still
definitively formulated \cite{boliok}. Probably, the most common
paradigm trying to include both DE and DM, is the so-called
$\Lambda$CDM model \cite{cop}. In this model, a cosmological
constant term, $\Lambda$, is included within the Einstein equations
and the total matter density is assumed to be the sum of baryonic
and DM densities, i.e. $\rho=\rho_b+\rho_{DM}$. This model fits
excellently and with high-precision all observational data,
meanwhile it provides a remarkably small number of cosmological
parameters \cite{cop2}. The effects of $\Lambda$ are \emph{hidden}
in the definitions of the pressure, $p$, and of matter density,
$\rho$, i.e. $p\rightarrow p+\frac{\Lambda c^2}{8\pi G}$,
$\rho\rightarrow\rho-\frac{\Lambda c^2}{8\pi G}$; thus, by assuming
hereafter a (flat) Friedman-Robertson-Walker (FRW) metrics, i.e.
$ds^2=c^dt^2-a(t)^2\big[dr^2/(1-kr^2)+\sin^{2}\theta
d\phi^2+d\theta^2\big]$, we can easily write down the correspondent
Friedmann equations:
\begin{eqnarray}\label{ave2}
H^2&\equiv&\left(\frac{\dot a}{a}\right)^2={8\pi G\over3}\rho-\frac{kc^2}{R_{0}^{2}a^2}\,,\\
\frac{\ddot a}{a}&=&-{4\pi
G\over3}\left(3p/c^2+\rho\right)\,,\nonumber
\end{eqnarray}
where $G$ is the gravitational constant, and $R_0$ the radius of the
universe. If $p$ and $\rho$ are known (or equivalently $H$ is
known), the above coupled equations describe the dynamics of the
universe.

In spite of its simplicity and its triumph in fitting data, the $\Lambda$CDM
model suffers from several shortcomings \cite{ell}, as e.g. the
problem of fine tuning and the problem of coincidence. The problem
of fine tuning is connected to the large difference between the
value of vacuum energy in quantum field theory and the observed
value of the cosmological constant \cite{wei}. The problem of
coincidence is related to the incredibly small differences between
the densities of DE and DM, which are supposed to evolve differently
as the universe expands. 
Moreover,
one can notice that at large enough energies (typically of the order
of the Planck scale, $M_{pl} \equiv \sqrt{\hbar c/G}\sim 1.22\times
10^{19}$GeV/$c^{2}$), the $\Lambda$CDM model is supposed to be not
reliable anymore; in other words, it only provides a limited
description \cite{yup} of the early stages of the universe. As a
consequence, the crucial role played by inflation still remains an
effective approach without any basis in a fundamental theory.
Nevertheless, a Quantum Gravity (QG) theory seems to be necessary to
explain these stages \cite{cop}, having as a limit the
$\Lambda$CDM model; that is clear if one believes that GR is a
particular case of a more fundamental theory. This is the underlying
philosophy of what are usually referred to as modified theories of
gravity, concerning the modified Einstein-Hilbert (EH) action
\cite{fre,fre2}. Besides fundamental physics motivations,
cosmologists acquired a huge interest in all of these theories, thanks
to the possibility to reach a unified scheme. Among the various
alternatives, crucial advances have been carried out in the studies
of Black Hole theory and String theory \cite{fre3}. 

A very intriguing concept, within the framework of GR, is the so-called
\emph{holographic principle} (HP), which provides some clues for
solving the modern theoretical and observational problems without
directly modifying the EH action. As it has been pointed out in QG, the entropy 
of a system does not scale with the volume of that system, but with the area of its surface \cite{hol}. Starting
from the above consideration, the HP postulates that the maximum
entropy inside a region is not extensive, but grows as the area of
the surface. Therefore, the total number of independent degrees of
freedom should scale with the surface area (in Planck units) as
well. In particular, the principle invokes that $L^3\rho_{vac}\leq
LM_{Pl}^{2}$, where $M_{Pl}$ is the Planck mass and $\rho_{vac}$ is
the vacuum energy density of a system of size $L$. Actually, the HP
represents a new basic principle for both QG and GR, being supported
by an effective quantum field theory. In order to include this
principle in a real cosmological scenario, one needs to choose the
correct cosmological length scale $L$, which is not \emph{a priori}
known.

Since the cosmological system is the universe, the DE should be
associated with the scale density, namely
$\rho_{DE}\equiv\rho_{X}\propto L$. Writing the fraction of DE
density in the form $\Omega_{X}\equiv
\frac{\rho_{X}}{3M_{Pl}^2H^2}$, we can imagine different holographic
approaches by choosing different IR cutoffs. For instance, for the
case of the $\Lambda$CDM, if the Hubble parameter $H$ is taken to be
the characteristic scale of the universe, then it is natural to
postulate $\Lambda^{-1/2}\propto H$; unfortunately this scenario
fails in reproducing the positive acceleration of universe.
Nevertheless, from the HP principle, it would be possible to solve
both the problems of fine tuning and coincidence just by introducing
the correct length scale. In order to fix such a scale, many
approaches have been investigated in literature. In the so-called
{\it Holographic Dark Energy model} (HDE) \cite{burp1hde,burp1hde2}
the future event horizon of the universe characterizes the length
$L$, i.e.
\begin{equation}\label{un}
L \equiv a\int_{a}^{\infty}\frac{da^{\prime}}{Ha^{\prime2}}\,.
\end{equation}
An alternative is the so-called agegraphic model (ADE)
\cite{burp1hde3,burp1hde4,burp2ade5}, where the IR cutoff is the conformal age of the
universe,
\begin{equation}\label{du}
L\equiv \eta =\int \frac{da^{\prime}}{a^{\prime2}H}\,.
\end{equation}
Unfortunately, the choice of a length scale is neither so easy nor
arbitrary, since it leaves unclear some conceptual problems. As pointed out by Cai
\cite{cai}, a drawback, concerning causality, appears in
the above scenarios: how is it possible that a local quantity, as
e.g. the DE, could be explained by global concepts, derived from the
physics of space-time? Is there a mechanism allowing a local
quantity to be determined by a global one \cite{nuevo}? A crucial
issue is that the above cosmological lengths are originated from an
expanding universe, which is an \emph{a priori} assumption and not
the result of a certain model, as it should be. In order to solve
this problem and inspired by the HP, we suggest to choose as length
scales only those quantities which are invariant under geometrical
transformations, avoiding the causality issue. This allows to solve
both the coincidence and fine-tuning problems, as well. The
invariants of curvature, derived from the Ricci scalar $R$, the Weyl
tensor $C_{\mu\nu\xi\delta}$ and the Riemann tensor
$R_{\mu\nu\xi\delta}$, are based on such theoretical
implicit assumptions and represent geometrical invariants. Gao \cite{gao} suggested, for the first time, 
that the DE density could be proportional to the first order invariant $R$:\footnote{Usually the
model is referred to as the Ricci Dark Energy (RDE) model.}
\begin{equation}\label{gao}
\rho_X\propto R\,.
\end{equation}
In this work we extend the analysis by Gao to the case of \emph{second
order} (independent) invariants. Notice that a second order
invariant is proportional to the inverse fourth power of an IR
cutoff, then indicating with $I_i$ the i-th independent second
order invariant, our assumption should be formulated as follows
\begin{equation}\label{postu}
\rho_X\propto\sqrt{|I_i|}\,.
\end{equation}
The purpose of the present article is to study the models
originating from these invariants and their implications on the FRW
spacetime. As shown below, these models fairly overcome the problem of the length
scale, providing a good solution for the DE
problem \cite{cop} and indicating a good agreement with the theoretical
predictions.

The paper is organized as follows: in Section II we develop the
theory of the independent second order invariants for a flat FRW
universe and we analyze the outcomes of the assumption
($\ref{postu}$) in cosmology. In Section III we fit our models with
the cosmological data from Supernovae Ia (SNeIa), Baryonic Acoustic
Oscillation (BAO) and Cosmological Microwave Background (CMB). In
Section IV, we provide a comparison with other relevant cosmological
models and we make use of model independent statistical tests, i.e.
the AIC and BIC. Finally in Section V we present the conclusions.

\section{Second order holographic invariants}
\label{sec2}

The basic purpose of this work
is to invoke the HP and to relate the second order curvature
(independent) invariants to the DE density (as already discussed,
the case of the first order curvature invariant has been extensively
treated by Gao \cite{gao,gao2,gao3}).

This can be easily performed by writing down the explicit form of
the second order geometrical invariants for a FRW metric and by
using eq. ($\ref{postu}$). As pointed out in Ref.\cite{capozzox}, 
among the 14 curvature scalar invariants \cite{stephani}, the most
interesting ones are the Kretschmann, the Chern-Pontryagin and Euler
invariants. We can write, for every
spacetime \cite{gendeb,Witten,Petrov,rinpen,carmi}, their
expressions as follows
\begin{eqnarray}
K_1 &=& R_{\alpha\beta\gamma\delta}R^{\alpha\beta\gamma\delta}\,, \nonumber \\
K_2 &=& [{}^*R{}]_{\alpha\beta\gamma\delta}R^{\alpha\beta\gamma\delta}\,, \nonumber \\
K_3 &=&
[{}^*R{}^*]{}_{\alpha\beta\gamma\delta}R^{\alpha\beta\gamma\delta}\,,
\end{eqnarray}
where the stars indicate the correspondent dual counterparts. From
the first Matt\'e-decomposition of the Weyl tensor, it is easy to
get \cite{roberts}
\begin{eqnarray}
R_{\alpha\beta\gamma\delta}&=&C_{\alpha\beta\gamma\delta}+\frac{1}{2}
\big( g_{\alpha \gamma} R_{\beta \delta}-\nonumber\\
&&-g_{\beta \gamma} R_{\alpha \delta}- g_{\alpha \delta}R_{\beta \gamma}+ g_{\beta \delta}R_{\alpha \gamma}\big)\nonumber \\
&&-\frac{1}{6}( g_{\alpha \gamma}g_{\beta \delta}-g_{\alpha
\delta}g_{\beta \gamma})R\,.
\end{eqnarray}
Therefore, $K_1$, $K_2$ and $K_3$ can be expressed as follows
\begin{eqnarray}\label{KAPPA1}
K_1&=&C_{\alpha\beta\gamma\delta}C^{\alpha\beta\gamma\delta}+2R_{\alpha\beta}R^{\alpha\beta}-\frac{1}{3}R^2=\, \nonumber  \\
&=&I_1 + 2R_{\alpha\beta}R^{\alpha\beta}-\frac{1}{3}R^2\, ,\\
K_2&=&[{}^*C{}]_{\alpha\beta\gamma\delta}C^{\alpha\beta\gamma\delta}=I_2\,\nonumber\\
K_3&=&-C_{\alpha\beta\gamma\delta}C^{\alpha\beta\gamma\delta}+2R_{\alpha\beta}R^{\alpha\beta}-\frac{2}{3}R^2= \nonumber \\
&=&-I_1+2R_{\alpha\beta}R^{\alpha\beta}-\frac{2}{3}R^2\,.\nonumber
\end{eqnarray}
From the above equations it is straightforward to infer the
explicit expressions of the second order invariants in a FRW
universe:
\begin{eqnarray}\label{primo}
I_1&=&\frac{60}{c^4}\left\{(\dot
H+2H^2)^2+H^4+\frac{2H^2kc^2}{R_0^2 a^2}+\frac{k^2c^4}{R_0^4 a^4}\right\}\,,\nonumber\\
I_2&=&0\,,\nonumber\\
I_3&=& -\frac{12}{c^4}\{5(\dot
H+2H^2)^2+5H^4+\frac{10kc^2H^2}{R_0^2 a^2}+\frac{5k^2c^4}{R_0^4 a^4}\nonumber\\
&+& 2(\dot H+2H^2)H^2+2(\dot H+2H^2)\frac{kc^2}{R_0^2 a^2}\}\,.
\end{eqnarray}
The HP postulate reads
\begin{equation}\label{cnn}
\rho_X=\frac{3\alpha}{8\pi G}\sqrt{|I_i|}\,,\quad i=1,2,3;
\end{equation}
where $\alpha$ is an dimensionless constant and should
not be confused with the tensorial index.

By combining together eqs. ($\ref{cnn}$) and ($\ref{ave2}$) and by
using the above expressions for $I_1$ and $I_3$ (the only two
non-trivial invariants), we obtain two differential equations, each
one providing the temporal evolution of the Hubble parameter. A
first interesting step consists in solving the associated
differential equations numerically for both $I_1$ and $I_3$, finding
the correspondent acceleration parameters, by using the definition
\begin{equation}\label{qu0}
q(t)=-\frac{\dot H}{H^2}-1\,,
\end{equation}
and the effective barotropic parameters, expressed by
\begin{equation}\label{wuz}
w(t)=-1-\frac{1}{3H}\frac{d}{dt}\ln\rho\,,
\end{equation}
for both the models.

\begin{figure}[h!]
\includegraphics*[scale=0.3]{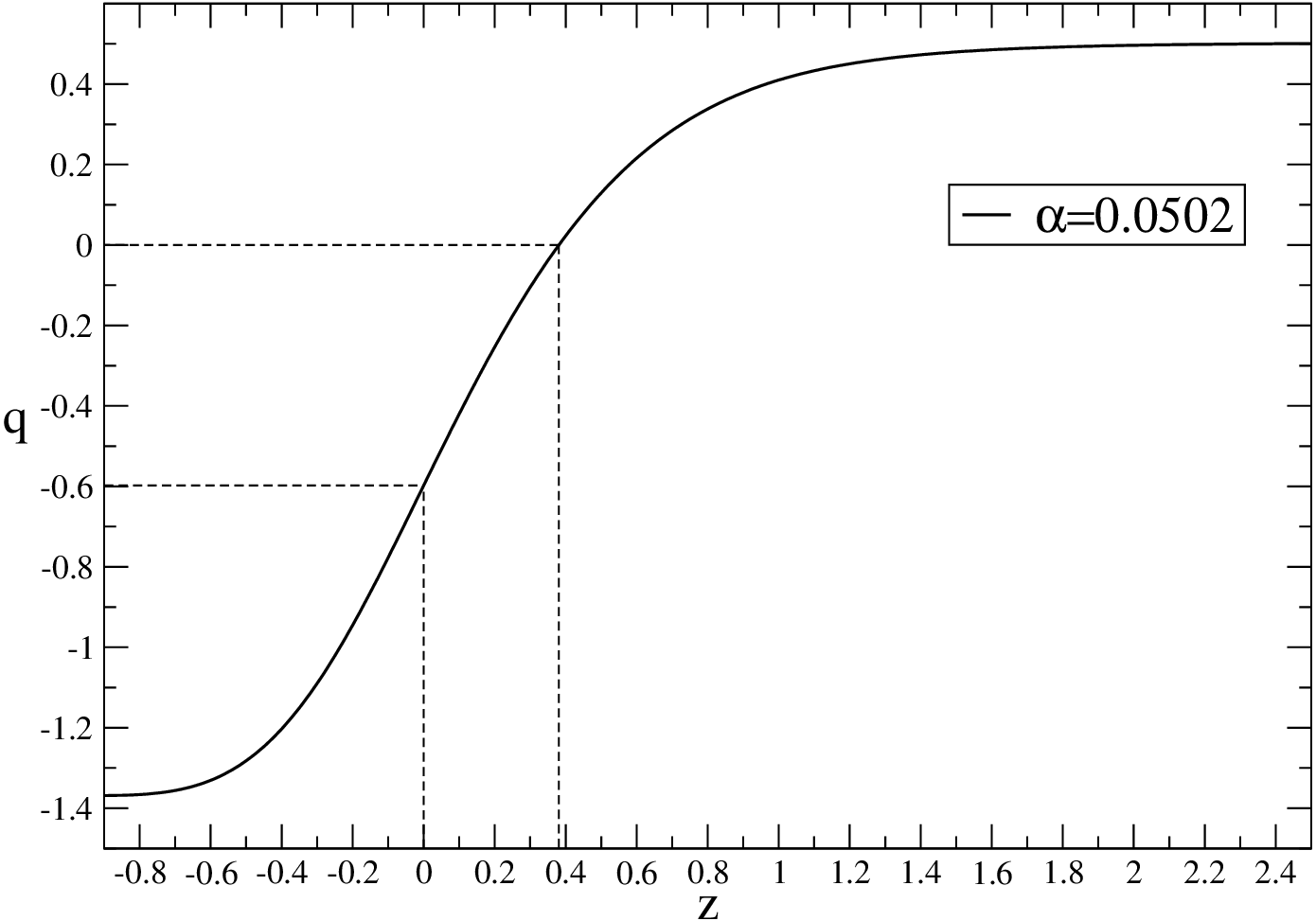}!
\caption{Deceleration parameter $q$ as a function of the redshift
$z$ for $mod_1$ (in this case we use conventionally $\alpha=0.050$).
Note that $q$ is negative for $z<0.4$, providing an accelerated
behavior of the universe since $z=0.4$; this is quite in agreement
with the prediction of $\Lambda$CDM which refers to $z\sim0.77$.
}\label{1q}
\end{figure}

\begin{figure}[h!]
\includegraphics*[scale=0.3]{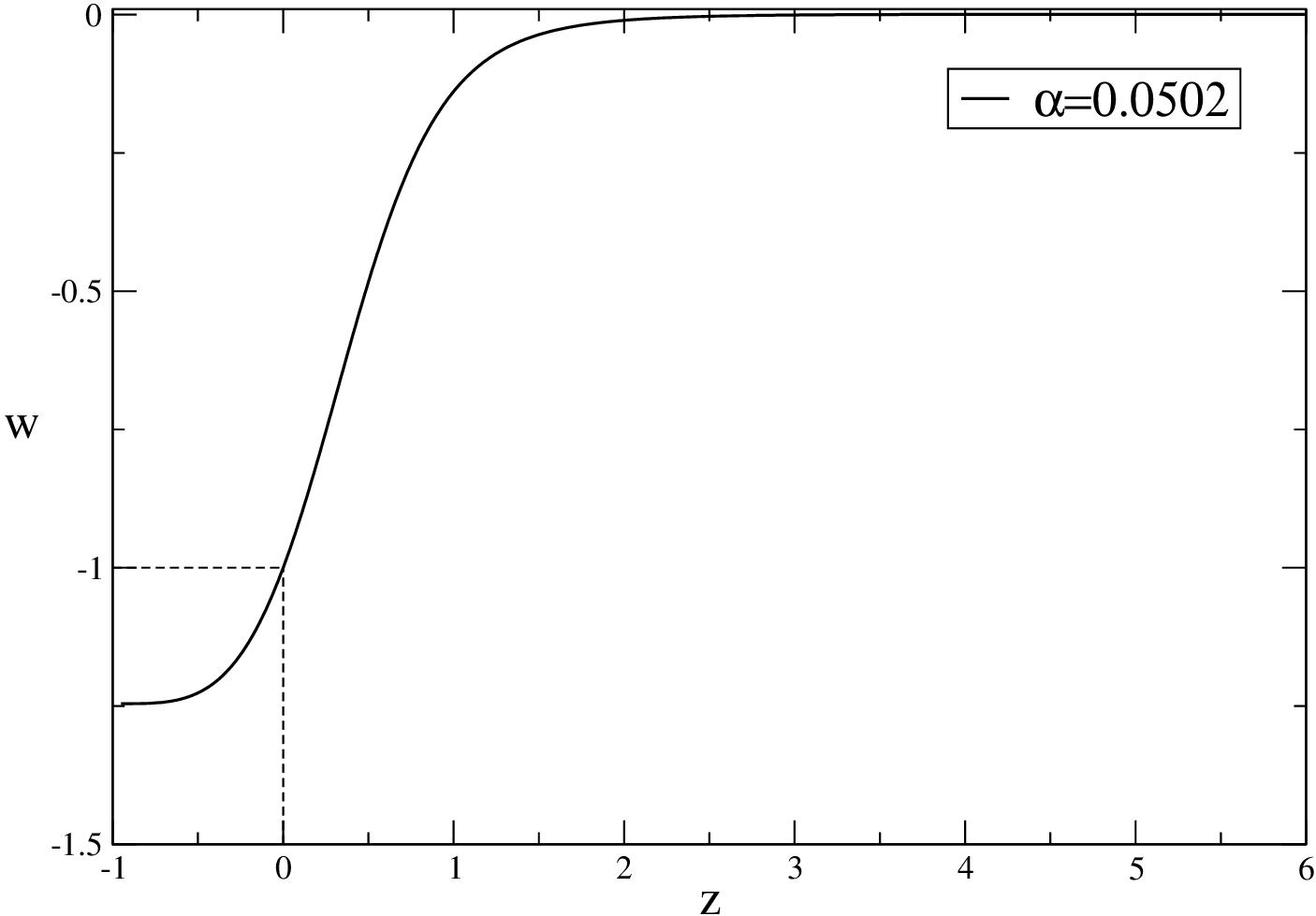}
\caption{The barotropic factor $w\equiv\frac{p}{\rho}$ as a function
of the redshift $z$ for $mod_1$ (in this case we use
$\alpha=0.050$). Notice that $w=-1$ at $z=0$, behaving as a
cosmological constant at low redshift.}\label{1w}.
\end{figure}

\begin{figure}[h!]
\includegraphics*[scale=0.3]{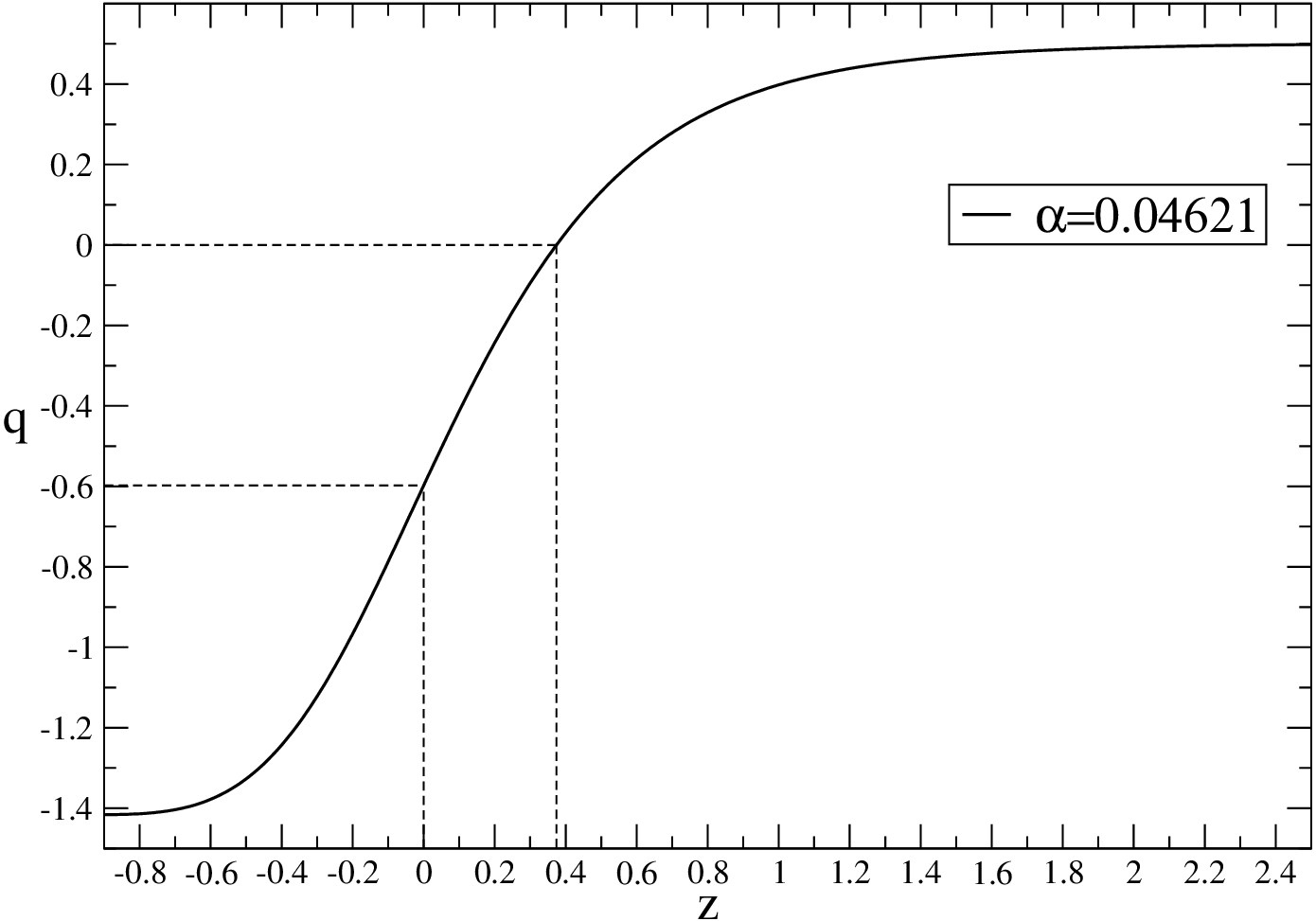}
\caption{Deceleration parameter $q$ as a function of the redshift
$z$ for $mod_3$ (in this case we use $\alpha=0.046$). Again $q$ is
negative for $z<0.4$, providing an accelerated behavior of the
universe since $z=0.4$ until now.}\label{3q}
\end{figure}

\begin{figure}[h!]
\includegraphics*[scale=0.3]{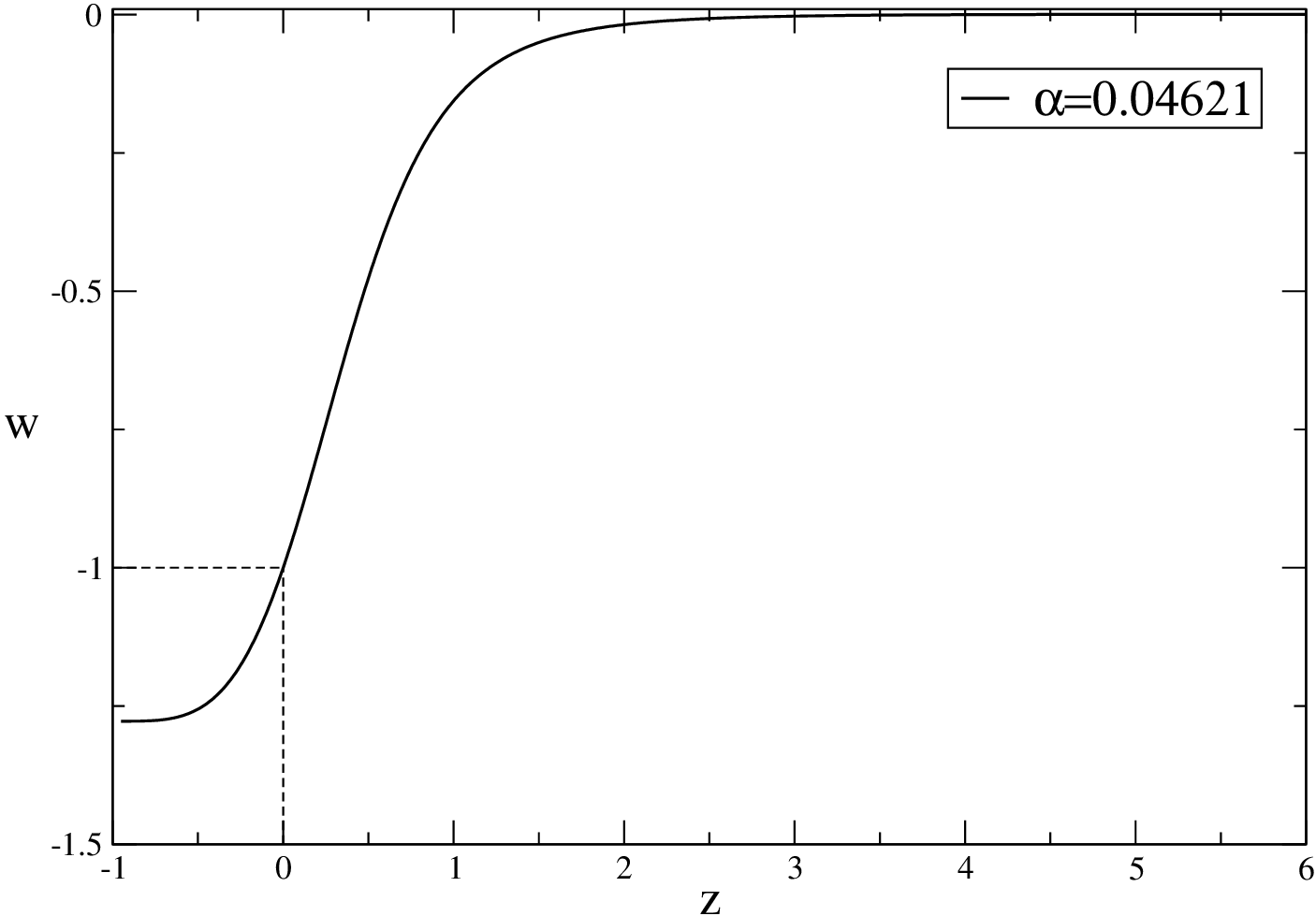}
\caption{The barotropic factor $w$ as a function of the redshift $z$
for $mod_3$ (in this case we use $\alpha=0.046$). Notice that again
$w=-1$ at $z=0$, behaving in this case too as a cosmological
constant term at low redshift.}\label{ew}
\end{figure}

From now on, we will refer to the cosmological models arising from
the invariants $I_1$ and $I_3$ as $mod_1$ and $mod_3$, respectively.
In order to fix the free parameters of our two models, we need to
find for which values of $\alpha$, the conditions $q_0<0$ and $w\sim-1$ at redshift $z=0$ are fulfilled.
The corresponding intervals are $\alpha\in[0.042,0.052]$ and $\alpha\in[0.039,0.047]$ for
$mod_1$ and $mod_3$ respectively. In figures \ref{1q}, \ref{1w}, \ref{3q} and \ref{ew} we
plot $q(z)$ and $w(z)$ as functions of the redshift for both $mod_1$
and $mod_3$. As shown in the pictures, if we set $\alpha=0.050$ for
$mod_1$ and $\alpha=0.046$ for $mod_3$ we get the expected
accelerated behavior of the universe, being $q\sim-0.6$ and $w\sim-1$. In particular,
the results obtained with the two models suggest that the universe
is accelerating since $z\sim 0.4$.

Once determined $q(z)$ and $w(z)$ we can characterize completely
the kinematics of the models. In the following section we will check
the agreement of our models with the observations.
In particular, we will make use of three independent tests: SNeIa,
BAO and CMB.

\section{Cosmological constraints}

In this section we perform an experimental
combined procedure, by using a fairly typical combination of
kinematical data. In particular we employ
the three most common fitting procedures: SNeIa, BAO and
CMB; the first two concern low redshift data sets
spanning from $z=0$ to $z\sim2$, while the third is a higher
redshift test, since it is performed at $z\sim1000$.

It is well established that standard candle data from SNeIa are
indicators of distance, able to fit the correspondent luminosity
distances for a particular class of models, by considering the
distance modulus and the correspondent redshift $z$. The
problem of the systematics, which usually affects these measures,
can be avoided or at least reduced, by the use of the most recent
updated Union 2 compilation \cite{kow}, instead of other older
samples \cite{oldkow}. Then, associating to each Supernova
modulus $\mu$ the corresponding $1\sigma$ error, denoted by
$\sigma_{\mu}$, we can perform directly the experimental analysis.
To this purpose, let us rewrite the distance modulus
\begin{equation}
\mu = 25 + 5 \log_{10} \frac{d_l}{Mpc}\,,
\end{equation}
where $d_L(z)$ is the luminosity distance, defined by
\begin{equation}
d_L(z) =  c (1+z) \int_0^z \frac{dz'}{H(z')}\,.
\end{equation}
Since the likelihood function $\mathcal{L}$ is related to the
chi-square statistic, i.e. $\mathcal{L}\propto \exp(-\chi^{2}/2)$,
we constrain the free parameters of a model, by minimizing the
quantity
\begin{equation}
\chi^{2}_{SN} =
\sum_{i}\frac{(\mu_{i}^{\mathrm{theor}}-\mu_{i}^{\mathrm{obs}})^{2}}
{\sigma_{i}^{2}}\,.
\end{equation}

The second test that we perform is related to the observations of
large scale galaxy clusterings, which provide the signatures of the
BAO \cite{Percival:2009xn}. In particular we use the measurement of
the peak of luminous red galaxies observed in Sloan Digital Sky
Survey (SDSS), usually denoted by $A$ and defined as follows

\begin{equation}
A=\sqrt{\Omega_m}  \big[\frac{H_0}{H(z_{BAO})}\big]^{\frac{1}{3}}
\left[ \frac{1}{z_{BAO}}\int_0^{z_{BAO}}
\frac{H_0}{H(z)}dz\right]^{\frac{2}{3}}\,,
\end{equation}

with $z_{BAO}=0.35$. In addition, the observed $A$ is estimated to
be
\begin{equation}\label{ax}
A_{obs} = 0.469 \left(\frac{0.95}{0.98}\right)^{-0.35}\,,
\end{equation}
with an error $\sigma_A = 0.017$. In the case of the BAO measurement
we should perform the following minimization:
\begin{equation}
\chi^{2}_{BAO}=\left(\frac{A-A_{obs}}{\sigma_A}\right)^2\,.
\end{equation}

Finally, concerning the CMB, we first analyze the so-called CMB
shift parameter
\begin{equation}
R=\sqrt{\Omega_m} \int_0^{z_{CMB}} \frac{H_0}{H(z)} dz\,.
\end{equation}

Since this standard ruler is fixed by the sound horizon at
decoupling ($z_{\rm dec}=1091.36$ \cite{Wang:2007mza}), it gives a
complementary bound to the SNeIa data and BAO as well. For the $R$
parameter, the observed value is $R_{obs} = 1.726\pm0.018$, as
inferred from the WMAP 7 data \cite{kow}. Hence, the correspondent
CMB constraints are given by minimizing the chi square
\begin{equation}
\chi^{2}_{CMB}=\left(\frac{R-R_{obs}}{\sigma_R}\right)^2\,.
\end{equation}

Moreover, we notice that, differently from SNeIa, BAO and CMB do not depend on
$H_0$. 

Finally, by combining the minimization procedures for SNeIa,
CMB and BAO, we constrain the parameters of our two models. 
The numerical results are summarized in the following
tables:
\begin{table}[h]
\begin{center}
\begin{tabular}{|c|c|c|c|}
\hline
$\Omega_{m}(SN)$  &  $\Omega_{m}(BAO)$ & $\Omega_{m}(CMB)$ \\
\hline
$0.240\pm0.070$ & $0.234\pm0.014$ & $0.266\pm0.064$ \\
\hline
$\alpha(SN)$  &  $\alpha(BAO)$ & $\alpha(CMB)$ \\
\hline
$0.046\pm0.016$ & $0.044\pm0.010$ & $0.025\pm0.016$\\
\hline
\end{tabular}
\caption{Summary of the results for $mod_1$; in this case we have
$\chi^{2}_{SNeIa}=1.021$, $\chi^{2}_{BAO}=1.001$,
$\chi^{2}_{CMB}=1.001$, $\Omega_{m,(mean)}=0.247\pm0.070$,
$\alpha_{(mean)}=0.038\pm0.016$.} \label{tabella1}
\end{center}
\end{table}

\begin{table}[h]
\begin{center}
\begin{tabular}{|c|c|c|c|}
\hline
$\Omega_{m}(SN)$  &  $\Omega_{m}(BAO)$ & $\Omega_{m}(CMB)$ \\
\hline
$0.260\pm0.090$ & $0.312\pm0.080$ & $0.343\pm0.060$ \\
\hline
$\alpha(SN)$  &  $\alpha(BAO)$ & $\alpha(CMB)$ \\
\hline
$0.042\pm0.010$ & $0.042\pm0.002$ & $0.024\pm0.0015$\\
\hline
\end{tabular}
\caption{Summary of the results for $mod_3$. The reduced chi squared
are $\chi^{2}_{SNeIa}=1.021$, $\chi^{2}_{BAO}=1.001$,
$\chi^{2}_{CMB}=1.003$; the mean values
$\Omega_{m,(mean)}=0.305\pm0.090$, $\alpha_{(mean)}=0.036\pm0.015$.}
\label{tabella2}
\end{center}
\end{table}

The results show that the theoretical predictions
discussed in Sec.\ref{sec2} are in agreement with the experimental
ones. The mean values of $\alpha$ in the tables are indeed included in the intervals of values 
found in Sec.\ref{sec2}. The CMB measurements are usually smaller if compared with
the SNeIa and BAO ones. Notice finally that for the SNeIa we had to fix a value for $H_0$ ($H_0=2.33\time10^{-18}\,s^{-1}$), while for CMB and BAO it is not necessary.

\section{comparison with other models}

In the Introduction we pointed out that the $\Lambda$CDM paradigm appears
to be the favorite fitting model among
a large number of possibilities. This should depend on its small
number of parameters. In particular, for a flat
cosmology, the only parameter involved is the mass density
$\Omega_m$.

Therefore, one can ask if there is a real necessity to go beyond
this approach, by considering other frameworks. This question
suggests the requirement to find a test able to compare different
cosmological models, in order to select the "best" one. At the same
time, such a test should also be model independent.

A good choice is represented by the so-called Akaike Information
Criterium (AIC) and BIC \cite{AIC,trs} tests, which are two of the
most model-independent statistical methods for comparing different
models. Moreover, since their first use by Liddle \cite{trn}, these
tests became a standard diagnostic
tool \cite{do,qua,quu} of regression models \cite{tro,qud,qut,ulp}.

The idea behind AIC is based on postulating two distribution
functions, namely $f(x)$ and $g(x|\theta)$: $f(x)$ is assumed to
be the exact one, while $g(x|\theta)$ approximates the former
through a set of parameters denoted by $\theta$. Hence, once given
$f(x)$ and $g(x|\theta)$, there exists only a set of $\theta_{min}$
minimizing the difference between $g(x,\theta)$ and $f(x)$
\cite{quqq}.

However, without going into details, we only note that the AIC value
for a single model is meaningless since the exact model function
$f(x)$ is unknown. Therefore, the quantities of interest are the
differences $\Delta AIC \equiv AIC - AIC_{min}$, calculated over the
whole set of models. The generic $AIC$ is given by
\begin{equation}
{\rm AIC}=-2\ln{\cal L}_{max}+2\kappa\,.
\end{equation}
A very similar criterion was derived by Schwarz \cite{qud} in a
Bayesian context (see \cite{qut} and references therein).

The BIC test provides
\begin{equation}
{\rm BIC}=-2\ln{\cal L}_{max}+k\ln N\,,
\end{equation}
where, for both AIC and BIC, ${\cal L}_{max}$ is the maximum
likelihood, $k$ is the number of parameters, and $N$ is the number
of data points used in the fit. If the errors are Gaussian, then
$\chi_{min}^2=-2\ln{\cal L}_{max}$ and the difference in BIC can be
simplified to $\Delta{\rm BIC}=\Delta\chi_{min}^2+\Delta k\ln N$.

We adopted both the AIC and BIC tests for different models. 

A first natural extension of the $\Lambda$CDM paradigm is the wCDM model,
also called quintessence model; it provides an EoS of the
form $p=w\rho$ with a negative barotropic factor, whose origin is related to the coupling of the Ricci scalar
with a not evolving scalar field $\phi$. Since this model
is strongly dependent on $w$, it is called $w$CDM in analogy with
the $\Lambda$CDM.

The theoretical explanation about the origin of the scalar field
giving a negative EoS remains unsolved in the $wCDM$, then 
a varying quintessence has been proposed: if $w$
evolves with the redshift $z$, i.e. $w=w[z]$, the scalar field
evolves as well. So the origin of the latter should be found
thermodynamically or in other ways. 

One of the more recent and
intriguing varying quintessence model has been proposed by the
Chevallier, Polarsky, Linder (CPL) \cite{CPL}. This parametrization
suggests that $w[a]=w_0+w_1(1-a)$. Assuming
$a\equiv(1+z)^{-1}$, one gets $w[z]= w_0 + w_1 \frac{z}{1+z}$, which
for low and very high redshifts becomes constant, i.e.
$w(z\rightarrow0)=w_0$ and $w[z\rightarrow\infty]=w_0+w_1$.

We perform the AIC and BIC tests for our two models ($mod_1$ and
$mod_3$) and for the models discussed above. Moreover, we also
include the Ricci DE model (RDE), studied by Gao \cite{gao}. The
normalized Hubble rates, $E\equiv\frac{H}{H_0}$, for these models
are written below (notice that there is not an analytic expression
of $E$ for the models $mod_1$ and $mod_3$):
\begin{eqnarray}\label{ciao}
E_{\Lambda CDM}&=&\sqrt{\Omega_m(z)+1-\Omega_m}\,,\nonumber\\
\,\nonumber\\
E_{wCDM}&=&\sqrt{\Omega_m(z)+(1-\Omega_m)(1+z)^{3(1+w)}}\,,\nonumber\\
\,\\
E_{CPL}&=&\sqrt{\Omega_m(z)+\left(1-\Omega_m\right)f(z)}\,,\nonumber
\,\nonumber\\
\,\nonumber\\
E_{RDE}&=&\sqrt{\frac{2}{2-\alpha}\Omega_m(z)+\Omega_{f}(1+z)^{4-\frac{2}{\alpha}}}\,,\nonumber
\end{eqnarray}
where $f(z)=(1+z)^{3(1+w_0+w_1)}\exp\left\{-\frac{3 w_1
z}{1+z}\right\}$ and $\Omega_m(z)\equiv\Omega_m(1+z)^3$.

The results of the tests are summarized in the table below, in
which we report, for each model, the number and the names of
the free parameters, the $\chi^2$ and the values of BIC and AIC:

\begin{table}[h]
\begin{center}
\begin{tabular}{|c|c|c|c|c|c|}
\hline
Model & N. Par. (k-1)  &  Param. & $\chi^2_{min}$ & $\Delta BIC$ & $\Delta AIC$\\
\hline
$\Lambda CDM$ & $1$ & $\Omega_m$ & $557.40$ & $0$ & $0$ \\
\hline
Quint. & $2$ & $\Omega_m , w$ & $557.28$ & $6.20$ &$1.88$\\
\hline
CPL & $3$ & $\Omega_m , w_0,w_a$ & $557.45$ & $12.69$ & $4.05$\\
\hline
RDE & $2$ & $\Omega_m , \alpha$ & $573.72$ & $22.64$ & $18.32$\\
\hline
$mod_1$  & $2$ & $\Omega_m , \alpha$ & $568.86 $ & $17.79$ & $13.46$\\
\hline
$mod_3$  & $2$ & $\Omega_m , \alpha$ & $568.91$ & $17.84$ & $13.51$\\
\hline
\end{tabular}
\end{center}
\end{table}

where we have used: $\Omega_m
= 0.254 \pm 0.038$ for $\Lambda$CDM; $\Omega_m = 0.325 \pm 0.049$
and $w = -1.18 \pm 0.18$ for $w$CDM; $\Omega_m = 0.246 \pm 0.034$,
$w_0 = -0.91 \pm 0.12$ and $w_a = -0.32 \pm 0.04$ for CPL
\cite{hlde,linder}; $\Omega = 0.310 \pm 0.052$ and $\alpha = 0.380
\pm 0.049$ for the Ricci DE. 

The results have been obtained through a
direct analysis of each model, following the
combined procedure explained in Section III. We can conclude that
our models are disfavored by the AIC and BIC analysis if compared
with $\Lambda$CDM and $w$CDM. It is clear that $\Lambda$CDM remains
the favorite model. On the other hand, the results obtained for the
Ricci DE are even worst. This seems to suggest that using higher
order invariants is better than using invariants of lower order,
encouraging further investigations in our direction. Concerning the CPL
model, it is interesting to notice that, to a slight variation of
$w_0$ and $w_a$, it corresponds a large variation of AIC and BIC.
This is in agreement with the fact that CPL is a three
parameters approach, then it appears disfavored if compared with
$mod_{1}$ and $mod_3$.

\section{conclusion}

In this work we have investigated the possibility to relate the DE
term with the second order independent invariants $I_i$. This has
been performed in analogy with the work of Gao \cite{gao}, who
proposed a first order invariant approach, namely the Ricci DE. The
validity of our idea is based on the holographic principle, which
requires the existence of a cut-off scale, characterizing the dark
energy. Moreover, the latter must scale with the inverse square of
the cut-off (which means $\rho_X\propto\sqrt{I_i}$). After solving
numerically the Friedmann equations for the two non-trivial
invariants in a flat FRW universe, we tested the two resulting
models, namely $mod_1$ and $mod_3$. Hence, we first analyzed the
kinematics of the two models in terms of the acceleration parameter
$q(z)$ and in terms of the barotropic factor $w(z)$, showing that
they are compatible with the conditions $q<0$ and $w\sim-1$ at
$z=0$. Then, we developed a test combining SNeIa, BAO and CMB in
order to fix the cosmological parameters; again in this case we
found a good agreement with the experimental results. It is also important 
to notice that the two models, arising from different invariants, show an incredibly similar
behavior.

A robust evidence of the validity of the models is given
by the AIC and BIC tests, where we get results better than those
obtained with the Ricci DE approach. Therefore, our models are
compatible with the holographic principle and seem to behave better
than the first order invariant proposed by Gao. This encourages to
study higher order invariants, what will be done in future
works.

\section*{Acknowledgements}

It is a pleasure to thank Prof. Luca Amendola, Dr. Andrea Geralico, 
Prof. Salvatore Capozziello and Prof. Hernando
Quevedofor very fruitful discussions.

\end{document}